\documentstyle[epsfig]{article}
\newcommand  \f  \varphi

\newcommand{\be}{\begin{equation}}
\newcommand{\ee}{\end{equation}}
\newcommand{\ben}{\begin{displaymath}}
\newcommand{\een}{\end{displaymath}}
\newcommand{\ba}{\begin{eqnarray}}
\newcommand{\ea}{\end{eqnarray}}
\newcommand{\ban}{\begin{eqnarray*}}
\newcommand{\ean}{\end{eqnarray*}}

\newcommand{\g}{\gamma}

\begin{document}
\begin{titlepage}
\rightline{\bf INFNFE-0998}
\vspace{1cm}

\begin{center}
{\huge \bf Sudakov effects in}
\\ \vspace{0.5cm}
{\huge \bf electroweak corrections}

\vspace{2cm}
{\Large
{\bf P. Ciafaloni$^{(a)}$ and  D. Comelli$^{(b)}$}}\\
\vspace{1cm}
{\it \large
(a) INFN sezione di Lecce, via Arnesano, 73100 Lecce \\
(b) INFN sezione di Ferrara, via Paradiso 12, 44100 
Ferrara}\\

\vspace{5cm}
{\large\bf Abstract}
\end{center}
\begin{quotation}

In perturbation theory
the  infrared structure of the electroweak interactions
produces large corrections proportional to
double logarithms $\log^2 \frac{s}{m^2}$ ,
similar to Sudakov logarithms in
QED,  when the scale $s$ is much larger than the typical mass $m$ 
of the particles running in the loops.
These energy growing corrections can be
particularly relevant for the planned Next Linear 
Colliders.
We study these effects  in the Standard Model 
for the process  $e^+\,e^-\rightarrow f\, \bar{f}$
and we compare them with similar corrections 
coming from   SUSY  loops.
\end{quotation}
\vspace{3cm}

\end{titlepage}

\def\baselinestretch{1.1}

\section{IR divergences: qualitative discussion}

Infrared (IR) divergences arise in perturbative calculations 
from regions of integration
over the momentum $k$ where $k$ is small compared to the typical scales of
the process. 
This is a well known fact in
QED for instance \cite{ Landau}  where  the problem of an
unphysical divergence is solved by giving the photon a fictitious mass
which acts a a cutoff for the IR divergent integral.
 When real (bremsstrahlung)
 and virtual contributions are summed, 
the dependence on this mass cancels and the
final result is finite \cite{ Landau}.
 The (double)
logarithms   coming from these 
contributions are large and, 
  growing with the scale,
 can spoil perturbation theory 
and need to be resumed. They are usually 
called {\rm Sudakov } double logarithms \cite{Sudakov}.
 In the case of electroweak corrections,
similar logarithms arise when the typical scale of the process
 considered is much larger than the mass of the particles 
running in the loops, 
typically the $W(Z)$ mass \cite{moulta,sirling,verza}. 
The expansion parameter results then
$\frac{\alpha}{4\sin^2\theta_w\pi}\log^2\frac{s}{M_W^2}$, 
which is already
 10 \%  for for energies $\sqrt{s}$ 
of the order of 1 TeV.
This kind of corrections  becomes therefore
particularly relevant for next generation of 
linear colliders (NLC \cite{NLC}).
In the case of corrections coming from loops with $W(Z)$s, 
there is no  equivalent of ``bremsstrahlung'' like in
QED or QCD: the $W(Z)$, unlike the photon,  has a definite nonzero mass and
is experimentally detected like a separate particle. In this way the full
dependence on the $W(Z)$ mass is retained in the corrections.
Other singularities arise in perturbation theory, 
namely those coming from the
ultraviolet (UV) region. These divergences can be treated with the usual
renormalization procedure and can be resummed through RGE equations. However
they produce single logs and we expect them to be asymptotically subdominant
with respect to the double logs of IR origin.

We consider here the process $e^+e^-\to f\bar{f}$ in the limit of
massless external fermions. Our notation is that $p_1$ ($p_2$) is the
momentum of the incoming $e^-$ ($e^+$) and  $p_3$ ($p_4$) is the
momentum of the outgoing $f$ ($\bar{f}$).
Furthermore, we define the Mandelstam variables: 
$s=(p_1+p_2)^2=2p_1p_2,
t=(p_1-p_3)^2=-\frac{s}{2}(1-\cos\theta),
u=(p_1-p_4)^2=-\frac{s}{2}(1+\cos\theta)$.
In the following we consider only 
the  dominant double logs corrections of IR and collinear origin coming from
one loop perturbation theory  and 
we neglect systematically single logs (IR, collinear or UV)
and ``finite'' contributions that do not grow with energy.
We discuss the kind of diagrams 
where we expect these corrections to be present\footnote{only
vertex and box corrections will be analyzed, since vacuum polarization
corrections give only single logs, both of ultraviolet and infrared origin.},
and evaluate them in the
asymptotic regime $s\gg M_w^2$.

\section{Sudakov logarithms in the vertices}

We will consider first as an example,  to have a grasp over the effect
of the IR double logs, 
the ``SM-like case'' in which a ``W boson'' having mass $M$ and
coupling with fermions like the photon is exchanged. We take the Born
QED amplitude as the reference tree level amplitude. Then we denote the tree
level photon exchange amplitude with ${\cal M}_0= i 
\frac{1}{s}e^2\bar{v}_e(p_1)
\g_\mu u_e(p_2)\bar{u}_f(p_3)\g_\mu v_f(p_4)$ and the tree level
 photon vertex
with ${\cal V}_0=-i e\bar{v}_e(p_1) \g_\mu u_e(p_2)$; $e$ is the
 electron charge.

Let us first consider IR divergences coming from vertex corrections.
Since we work in the limit of massless fermions, there is no coupling to the
Higgs sector. Moreover, by power counting arguments,
 it is easy to see that the
vertex correction where the trilinear gauge boson coupling appears is not IR
divergent. The only potentially IR divergent diagram is then the one of
fig. 1, where a gauge boson is exchanged in the t-channel. 
It is convenient to choose the  momentum of 
integration $k$ to be the one
 of the exchanged particle, the boson in this case. 
Then, by simple power
counting arguments it is easy to see that the IR divergence can only be
produced by regions of integration where $k\approx 0$.
The only potentially IR divergent
integral is then the scalar integral, 
usually called $C_0$ in the literature \cite{veltman}.
 Any other integral with $k_\mu,k_{\mu}k_{\nu}$ in the numerator
cannot, again
by power counting, be IR divergent.
To understand the origin of the divergences, let us consider the diagram
of fig.1
with all the masses  set to zero. 
For $k\approx 0$ the leading term of the vertex amplitude is given by:
\be\label{QED}
{\cal V}\approx 
-\frac{\alpha}{4 \pi}{\cal V}_0  \int\frac{d^4k}{i\pi^2}
\frac{(p_1p_2)}{k^2(kp_1)(kp_2)}
\approx
-\frac{\alpha}{2 \pi}{\cal V}_0 \int_0^1\frac{dx}{x}\int_0^{1-x}\frac{dy}{y}
\ee
We can see here the two logarithmic
divergences that arise from the integration over the $x,y$ Feynmann
parameters. As is well known \cite{Landau}, one of them is of collinear
origin and the other one is a proper IR divergence. When we take some of the
external squared momenta and/or masses different from zero, they serve as
cutoffs for the divergences. Let us consider now some simple cases 
that will be useful in the following, where the cutoff is 
given by a single scale M.
The behavior for $C_0$ in the asymptotic region 
$s\equiv (p_1+p_2)^2\gg M^2$ is as follows:
\ba
C_0(m_1,m_2,M,p_1^2,p_2^2,s)&\equiv & \frac{1}{i \pi^2}
\int\frac{d^4k}{[(k+p_1)^2-m_1^2][k^2-M^2][(k-p_2)^2-m_2^2]}
\\
Re\{C_0(0,0,M,0,0,0,s)\}&\to& \frac{1}{2s}\log^2(\frac{s}{M^2})
\quad\mbox{ for }s\gg M^2\label{C0m=0}
\\
Re\{C_0(M,M,M,0,0,0,s)\}&\to& \frac{1}{2s}\log^2(\frac{s}{M^2})
\quad\mbox{ for }s\gg M^2\label{C0m=M}
\ea
Then we can use (\ref{C0m=0}) for the vertex of fig. 1 
in the asymptotic region finding:
\be\label{smvertex}
{\cal V}\approx -\frac{\alpha}{4 \pi} {\cal V}_0 \log^2(\frac{s}{M^2})
\quad\mbox{ for }s\gg M^2
\ee
where we can see the double logarithm behavior of the vertex correction
for $s\gg M^2$. 

The dependence on the IR logs simply factorizes for the cross section:
\[
\sigma\propto  \frac{1}{s}\int_0^s\frac{dt}{s}
|{\cal M}_0|^2 [1-2\frac{\alpha}{4 \pi} \log^2\frac{s}{M^2}]
=\sigma_0[1-2\frac{ \alpha}{4 \pi} \log^2\frac{s}{M^2}]
\]

Now let us consider the ``susy-like'' case in which 
a fermion is exchanged  and a scalar couples to the external gauge boson
(fig 1). In supersymmetry the internal  fermion and scalar,
for instance a neutralino and a selectron,  
have masses of the
same order and can, for our purposes, be taken to have 
the same mass $M$.
In fact the distinction between the two
masses is irrelevant as long as they are of the same order, since
$\log(\frac{s}{m^2})\log(\frac{s}{M^2})=
\log^2(\frac{s}{M^2})+\log(\frac{M^2}{m^2})\log(\frac{s}{M^2})
\approx \log^2(\frac{s}{M^2})$ and 
 we are interested only in double logs (single logs  are neglected).
Expression (\ref{smvertex}) is in this case
substituted by:
\be
{\cal V}\approx -
\frac{\alpha}{4\pi}{\cal V}_0\int\frac{d^4k}{i\pi^2}
\frac{M^2}{k^2(kp_1)(kp_2)}
\stackrel{s\gg  M^2}{\longrightarrow}
-\frac{M^2}{s}\frac{\alpha}{4 \pi} {\cal V}_0 \log^2(\frac{s}{M^2})
\ee
where we still have the double log behavior
 coming from the
 integration over the region
$k\approx 0$ (remember that always $s\gg  M^2$).
In this case however, $2 p_1 p_2=s$ is
substituted by $M^2$ in the numerator, 
so that the we have $\log^2\frac{s}{M^2}\to
\frac{M^2}{s}\log^2\frac{s}{M^2}$.
In the end the double logarithm behavior is strongly suppressed
by a factor $\frac{M^2}{s}$ for the SUSY vertex with respect to the
SM case . This is due to the different couplings that appear in the vertex
corrections: fermion-gauge boson coupling in the ``SM like'' case and 
fermion-scalar in the ``susy like'' case. 
In the first case the coupling is,
at high energy, proportional to $p_\mu^i$ where $i$ is the label of
 the external
fermion the exchanged boson couples to. 
Then we have a factor $p_i\cdot p_j$
where $i$ and $j$ are the fermions connected by the exchanged
 boson. In the ``susy
like'' case where  scalars  and fermions  are
 exchanged, no such factor is present and  
$p_i\cdot p_j$ gets substituted by $M^2$, generally subdominant at high
energies.

\section{Sudakov logarithms in the boxes}

Let us consider the exchange of a vector boson of mass M in the s-channel
(see fig. 2). 
In the limit  $M\to 0$ and in the IR region, the amplitude is given by:
\be\label{SMbox}
{\cal M}\approx
\frac{\alpha}{4 \pi}
{\cal M}_0
\int \frac{d^4k}{i\pi^2}
\left\{
\frac{(p_1p_3)}{k^2(p_1k)(p_3k)}
+
\frac{(p_2p_4)}{k^2(p_2k)(p_4k)}
\right\}
\ee
As is shown schematically 
in fig. 2, the two terms in this equation come from two different
region of integration. When $k\approx 0$, then $(k+p_1+p_2)^2\approx s$ and we
can think the ``upper'' boson line to be shrunk, like shown in the figure. 
The mirror situation is $k+p_1+p_2\approx 0,k^2\approx s$. This makes evident
the fact that the IR structure of the box is the same of the vertex.
Expression (\ref{SMbox}) is identical with  (\ref{QED})  
but with the difference that
$2 p_1 p_2=s$ gets substituted by $-2 p_1p_3=-2 p_2 p_4=t$. So
for SM boxes we have an exchange of s and t variables with respect
to SM vertices. In the end for the box contribution in the IR region we can
write:
\be\label{asSMbox}
{\cal M}\approx-\frac{\alpha}{2 \pi}{\cal M}_0
\log^2(\frac{t}{M^2})
\ee
It must be stressed however that this expression is valid only in the
asymptotic region $t\gg  M^2$ where the double log behavior is generated, 
while we assumed $s\gg  M^2$. 

Let us now consider the ``susy like'' box where a scalar particle is exchanged
in the t-channel (see figure 3). In this case the amplitude is:
\be\label{SUSYbox}
{\cal M}\approx
\frac{\alpha}{4 \pi}
{{\cal M}_0}
\int \frac{d^4 k}{i\pi^2} \left\{
\frac{M^2}{k^2(p_1k)(p_3k)}
+
\frac{M^2}{k^2(p_2k)(p_4k)}
\right\}
\ee

Comparing eqs. (\ref{SMbox}) and (\ref{SUSYbox}) we note that the susy
amplitude has a factor $\frac{M^2}{t}$ with respect to the SM one.
In the IR region $t \gg  M^2$ we have, using (\ref{C0m=M}):
\be\label{asSUSYbox}
{\cal M}\approx \frac{\alpha}{4 \pi} {\cal M}_0
 \frac{M^2}{t}\log^2\frac{t}{M^2}
\ee
Care must be taken when we compute cross sections since, as noted
above, eqs. \ref{asSMbox}) and \ref{asSUSYbox}) are valid only when 
$t\gg M^2$. Let us then 
consider a region of the phase space from a certain fixed value of $t$ of
order $s$ on, let's say
 $-s<t<-\frac{s}{2}$. 
Then, if $s\gg M^2$,  
we can use the expressions valid for $t\gg M^2$. Neglecting unessential
factors, the leading box 
corrections to the tree level cross sections are given by:
\ban
SM&
\Delta\sigma\approx\frac{\alpha}{s}\int_{-s}^{-\frac{s}{2}}\frac{dt}{s}\log^2
\frac{t}{M^2}\approx\sigma_0\alpha\log^2\frac{s}{M^2}\\
SUSY&
\Delta\sigma\approx\frac{\alpha}{s}\int_{-s}^{-\frac{s}{2}}\frac{dt}{s}
\frac{M^2}{t}\log^2
\frac{t}{M^2}\approx\sigma_0{\alpha}\frac{M^2}{s}\log^2\frac{s}{M^2}
\ean
Again, SUSY boxes are depressed by a power factor with respect to SM ones.

To conclude, we expect double logs of IR and collinear
 origin to give at high energies large
one loop corrections to observables in the SM.
 This is true both for box and
vertex corrections. On the other hand, in a susy theory, 
due to the different spins of the particles exchanged in the loops,
 these double logs are
expected to be power suppressed. For this reason, in the following we will
 consider in detail only SM electroweak corrections.

\section{Sudakov logarithms in the Standard model }

We study 
the purely electroweak  double logarithmic corrections
in the Standard Model 
coming from  the exchange of the $W$  and $Z$ gauge bosons
to the process $e^+\, e^-\rightarrow \bar{f}\,f$ in the massless case.

For the moment we consider only the massless external fermions 
$\mu $  for leptons, $u,\;c$ and $ d,\;s$   for  quarks,
and we neglect, for the moment, the bottom quark
whose  corrections contain a non trivial  flavor  dependence 
on the top 
mass ( future analyses ).

This kind of contributions, as explained before,  come from only 
vertex corrections  in which  one
gauge boson  it is exchanged
 and from the boxes   (direct and crossed)
with two $Z$s or two $W$s.


The  effective vertices 
$\gamma \,(Z)\,\bar{f}\,f$ including tree level and dominant double logs  
are given by
$\;\;\;
\bar{v}_e(p_1)\gamma_{\mu} (V_{fL}^{\gamma\,(Z)} 
\,P_L+V_{fR}^{\gamma\,(Z)} \,P_R)u_e(p_2)\;\;\;$ with
\ba
V_{fL}^{\gamma}&=&ig\,s_W  Q_f (1- \frac{1}{16 \pi^2} \frac{g^2}{c_W^2}
g_{fL}^{ 2} \log^2\frac{s}{m_Z^2} 
 -\frac{1}{16 \pi^2} \frac{g^2}{2}
\frac{Q_{f'}}{Q_f}
 \log^2\frac{s}{m_W^2}  
)
\\
V_{fR}^{\gamma}&=&ig \, s_W 
 Q_f (1- \frac{1}{16 \pi^2} \frac{g^2}{c_W^2}
g_{fR}^{ 2} \log^2\frac{s}{m_Z^2})
\ea
and
\ba
V_{fL}^{Z}&=& i\frac{g}{c_W}  g_{fL}
 (1- \frac{1}{16 \pi^2} \frac{g^2}{c_W^2}
g_{fL}^{ 2} \log^2\frac{s}{m_Z^2}  
-\frac{1}{16 \pi^2} \frac{g^2}{2}
\frac{g_{f'L}}{g_{fL}}
 \log^2\frac{s}{m_W^2}  
)
\\
V_{fR}^{Z}&=&i\frac{g}{c_W} 
 g_{fR} (1- \frac{1}{16 \pi^2} \frac{g^2}{c_W^2}
g_{fR}^{ 2} \log^2\frac{s}{m_Z^2})
\ea
Here f  is the external fermion and f' its isospin partner. 
Moreover, $g_{f(f')R}= -Q_{f(f')} s_W^2  $ and 
$ g_{f(f')L}=T_3^{f(f')}-Q_{f(f')}  s_W^2$.

Defining 
\ba \label{chiral}
\bar{v}_e(p_1)\g_\mu P_{L,R}u_e(p_2)\bar{u}_f(p_3)\g_\mu P_{L,R}v_f(p_4)
\equiv P_{L,R}\otimes P_{L,R}
\ea
the corrections from  box  diagrams come  from 
 direct and crossed diagrams 
as a sum  of projected amplitudes on the left-right
 chiral basis: 
\ba \nonumber
 B_{LL}  \; \gamma_{\mu} P_L \otimes\gamma^{\mu} P_L+
   B_{LR} \; \gamma_{\mu} P_L \otimes\gamma^{\mu} P_R+\\ \nonumber
B_{RL} \; \gamma_{\mu} P_R \otimes\gamma^{\mu} P_L+
B_{RR} \; \gamma_{\mu} P_R \otimes\gamma^{\mu} P_R
\ea
where
\ba\nonumber
B_{LL}&=& \frac{i}{s}\frac{g^4}{8 \pi^2}
 ( \frac{g_{eL}^2 g_{fL}^2}{c_W^4}
(\log^2 \frac{s+t}{m_Z^2}
-\log^2 \frac{t}{m_Z^2})+\\\nonumber
&&\frac{1}{4} ( \theta_{2 f} \log^2 \frac{s+t}{m_W^2}- \theta_{1 f}
\log^2 \frac{t}{m_W^2}))
\\\nonumber
B_{LR}&=& \frac{i}{s}\frac{g^4}{8 \pi^2} \frac{g_{eL}^2 g_{fR}^2}{c_W^4}
(\log^2 \frac{s+t}{m_Z^2}-\log^2 \frac{t}{m_Z^2})
\\\nonumber
B_{RL}&=& \frac{i}{s}\frac{g^4}{8 \pi^2}  \frac{g_{eR}^2 g_{fL}^2}{c_W^4}
(\log^2 \frac{s+t}{m_Z^2}-\log^2 \frac{t}{m_Z^2})
\\\nonumber
B_{RR}&=& \frac{i}{s}\frac{g^4}{8 \pi^2}  \frac{g_{eR}^2 g_{fR}^2}{c_W^4}
(\log^2 \frac{s+t}{m_Z^2}-\log^2 \frac{t}{m_Z^2})
\ea
 with the above expressions  obtained in the limit
$s,t \gg  M_{Z,W}^2$  and

$ \theta_{1 f}=1$ for $f= \mu,\,d$ and zero otherwise;

$ \theta_{2 f}=1$ for $f= \nu,\,u$ and zero otherwise;

The positive double log contributions come from the
 crossed box, while the
negative  ones from the direct  diagrams.

It is clear that the interference between the two
amplitudes,  for the exchange of $Z$ bosons, 
leads to a depression of the full
 contribution due to the fact that
\ba
\log^2 \frac{s+t}{m_Z^2}-\log^2 \frac{t}{m_Z^2}=
2 \log \frac{s}{m_Z^2 } \log \frac{1+\cos \theta}{1-\cos \theta}+ {\rm finite}
\ea
where {\rm finite}  means contributions not increasing as $\log s$.
In such a   way  we lose the leading $\log^2 s$ factor and we  remain  
 with a single $\log$ that we neglect.
So in leading approximation, box  diagram contributions come  only 
from $W$  exchange.

To obtain the physical observables
 we must square the full amplitude:
\ba \label{amplitude}
M= M_{LL}  \; \gamma_{\mu} P_L \otimes\gamma^{\mu} P_L+
  M_{LR} \; \gamma_{\mu} P_L \otimes\gamma^{\mu} P_R+\\ \nonumber
M_{RL} \; \gamma_{\mu} P_R \otimes\gamma^{\mu} P_L+
M_{RR} \; \gamma_{\mu} P_R \otimes\gamma^{\mu} P_R
\ea
where
\ba
M_{LL}&=&\nonumber 
-\frac{i}{s}(V_{eL}^{\gamma}V_{fL}^{\gamma}
+V_{eL}^{Z}V_{fL}^{Z})+B_{LL}\\\nonumber 
M_{RL}&=&
-\frac{i}{s}(V_{eR}^{\gamma}V_{fL}^{\gamma}+
V_{eR}^{Z}V_{fL}^{Z})+B_{RL}\\\nonumber 
M_{LR}&=&
-\frac{i}{s}(V_{eL}^{\gamma}V_{fR}^{\gamma}+
V_{eL}^{Z}V_{fR}^{Z})+B_{LR}\\\nonumber 
M_{RR}&=&
-\frac{i}{s}(V_{eR}^{\gamma}V_{fR}^{\gamma}+
V_{eR}^{Z}V_{fR}^{Z})+B_{RR}
\ea
and compute the differential cross section
\ba \label{cross}
\frac{d \sigma}{d \Omega}&=& \frac{s}{256 \pi^2 } N^f_c
 [(|M_{LL}|^2+|M_{RR}|^2)
(1+\cos \theta)^2+
\\
&&(|M_{RL}|^2+|M_{LR}|^2)(1-\cos \theta)^2]
\ea
with $N^f_c=1(3)$ for final state leptons (quarks) 
 and  $-1+ 2 \frac{m_Z^2}{s}<\cos \theta <1- 2 \frac{m_Z^2}{s}$
to be consistent with the above approximations ($t\gg  - m_Z^2$).
In any case we can extend the integration region to the full
$\pm 1$ range without modifying the leading results.

\section{ Sudakov logs
in the cross section  and in the forward backward asymmetry for  
$e^+\,e^-\rightarrow f\, \bar{f}$ }

We define 
$\sigma_{B}$ and $\sigma_{T}$ respectively as
the tree level (Born) cross section
 and as the total cross section containing  only 
 the  one loop  double logarithms .
The explicit expressions  for different  fermionic final states
are given by:

\ba
\sigma_{T}/\sigma_{B}\;(e^+\,e^-\rightarrow \mu\, \bar{\mu})&=& 
1+(-1.345_{Box}+0.282)\; \alpha_W-0.330 \; \alpha_Z 
\\
\sigma_{T}/\sigma_{B}\;(e^+\,e^-\rightarrow u\, \bar{u})&=& 
1+(-2.139_{Box}+0.864)\; \alpha_W  -0.385\; \alpha_Z \\
\sigma_{T}/\sigma_{B}\;(e^+\,e^-\rightarrow d  \, \bar{d})&=& 
1+(-3.423_{Box} +1.807)\; \alpha_W-0.557\; \alpha_Z
\ea
where $\alpha_{W,Z}=\frac{g^2}{16 \,\pi^2} \log^2 \frac{s}{m_{W,Z}^2}\simeq
2.7\, 10^{-3} \log^2 \frac{s}{m_{W,Z}^2}  $.
With the  underline ``$Box$''  we give the contributions  coming 
 from box diagrams, the rest is  from vertex corrections.

 For the  forward-backward asymmetry
$A_{FB}(e^+\,e^-\rightarrow f\,
\bar{f})$
the analytic expressions are:
\ba
A_{FB}^{T}/A_{FB}^{B}\;(e^+\,e^-\rightarrow \mu\, \bar{\mu})\!\!&=&\!\! 
1+(-0.807_{Box}+0.770)\;\alpha_W -0.002\; \alpha_Z
 \\
A_{FB}^{T}/A_{FB}^{B}\;(e^+\,e^-\rightarrow u\, \bar{u})\!\!&=&\!\!
1+(-0.521_{Box}+0.454)\;\alpha_W-0.023\; \alpha_Z
  \\
A_{FB}^{T}/A_{FB}^{B}\;(e^+\,e^-\rightarrow d  \, \bar{d}) \!\!&=& \!\!
1+(-0.620_{Box}+0.508)\;\alpha_W-0.029 \; \alpha_Z
\ea

We see that already at $\sqrt{s}=1\,(0.5)\;$ TeV the
 parameter $ \alpha_{Z,W}\simeq \,
6\,(3)\,10^{-2}$  so that the above corrections can exceed the ten (six)
 percent for the cross sections and
a  resummation technique (which is under study) is needed.

In the limit $\alpha_Z \simeq\alpha_W $  we can summarize the above
results in:

\ba 
&&
\frac{\sigma_{T}}{\sigma_{B}}(\mu\, \bar{\mu}) \simeq 1- 1.39 \;\alpha_{Z,W};
\;\;\;\;\;
\frac{A_{FB}^{T}}{A_{FB}^{B}}(\mu\,\bar{\mu})\simeq 1-0.04\;  \alpha_{Z,W};
\\
&&
\frac{\sigma_{T}}{\sigma_{B}}(u\, \bar{u})\simeq 1-1.66\; \alpha_{Z,W};
\;\;\;\;\;
\frac{A_{FB}^{T}}{A_{FB}^{B}}(u\,\bar{u})\simeq 1-0.09\;   \alpha_{Z,W};
\\
&&
\frac{\sigma_{T}}{\sigma_{B}}(d\, \bar{d})\simeq 1-2.17\; \alpha_{Z,W};
\;\;\;\;
\frac{A_{FB}^{T}}{A_{FB}^{B}}(d \,\bar{d})\simeq 1-0.11 \;  \alpha_{Z,W};
\\
\ea

We can make several comments to these results:
\begin{itemize}

\item  $Z$ boson exchange is  negative (photon-like) 
in the vertex corrections:
 it decreases both left and right effective 
vertices. In the boxes, Z exchange contribution does not gives a double log
behavior due to a cancellation between direct and crossed diagrams.
\item     
$W$  boson exchange, due to his chiral structure, affects
  only the left gamma vertex proportionally to $-\frac{Q_{f'}}{Q_f}$ and the
left $Z$ vertex  to $-\frac{g_{f'L}}{g_{fL}}$, giving always contributions
that are positive with respect to the tree level values.
Also box  diagrams  are peculiar because they affect
only the left-left structure of the amplitude and they always give
a negative contribution.
\item In  $\sigma_{T}/\sigma_{B}$ box corrections are dominant (more than 
three times  the vertex ones). Since, as noted above, box corrections
are given only by $W$ exchange,
the e.w. Sudakov corrections are a peculiar signature of the left-left
structure  of the full amplitude.
\item In $ A_{FB}^{T}/A_{FB}^{B}$  
 $Z$ corrections almost cancel. $W$ contributions from vertex are
accidently almost
 equal  and  opposite  to the  box's  ones
 leaving a  negligible contribution.
As a result, the double logs relative effect is 
more than one order of magnitude  smaller 
 than  for the full cross sections.
\item The total effect  from virtual double logs is negative both for 
the cross sections and for the asymmetries.
\end{itemize}

\section{Conclusions}
We have investigated, in  one loop electroweak
corrections,  the  IR origin of 
double logs that we denote as
e.w. Sudakov corrections.
These Sudakov effects can be important 
for next generation of colliders running at TeV energies
since they grow with energy like the square of a logarithm.
In supersymmetric models, loops containing the supersymmetric partners 
of the usual particles do not have double log asymptotical behavior
 (i.e., the double
logs are present but power suppressed).
In the SM the e.w. Sudakov corrections
are present with a peculiar chiral structure due to $W$ boson exchange
dominance; it should be possible to test 
the different  chiral contributions  with colliders with polarized beams. 
In any case, already for  TeV machines, proper resummation of such  
large contributions seems to be needed;
in fact for the various cross sections
we find that contributions of order 5-8 $\%$  are present for the
planed 500 GeV $e^+ e^-$ NLC \cite{NLC}.
The corrections to the asymmetries considered in this paper, 
due to the  accidental 
cancellation between  box and vertices  contributions, 
are almost negligible (one order of magnitude smaller with
respect to the cross sections relative corrections).
Sudakov effects in other kind of asymmetries (for instance polarized
asymmetries) and in general in other observables, are currently under
study. 

\vspace{1.cm}
{\Large\bf Acknowledgments}

The authors are indebted to M. Ciafaloni and C. Verzegnassi  for clarifying
discussions. A special acknowledgement goes 
to F. Renard for discussions and a
check of some of the computations.

\begin{figure}[htb]\setlength{\unitlength}{1cm}
\begin{center}
\begin{picture}(10,5)
\put(-.5,0){\epsfig{file=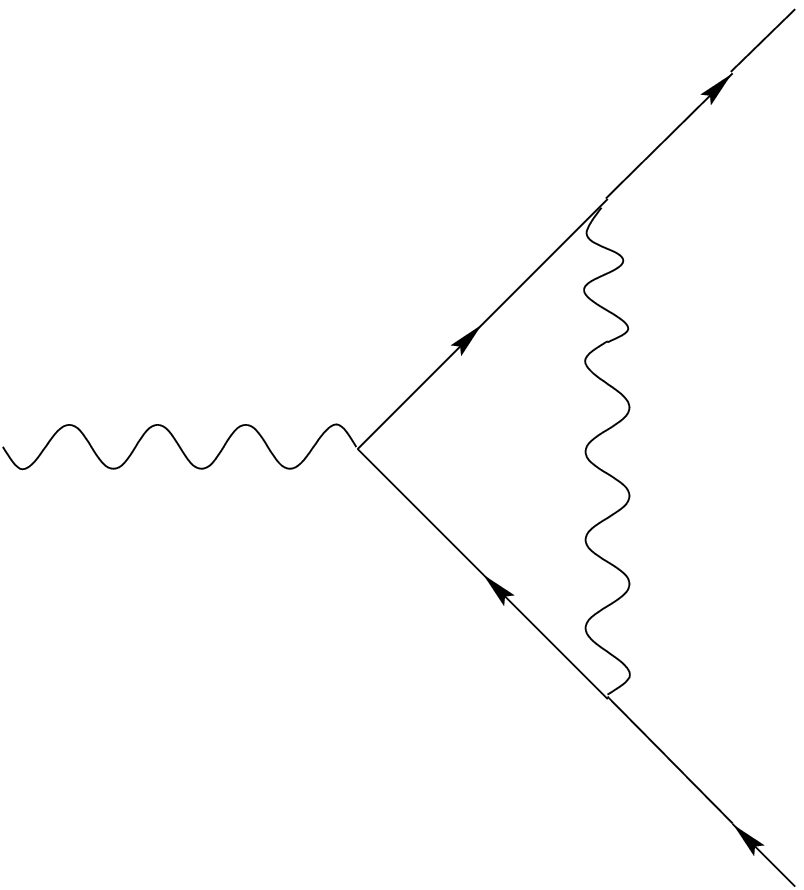,width=3cm}}
\put(7.5,0){\epsfig{file=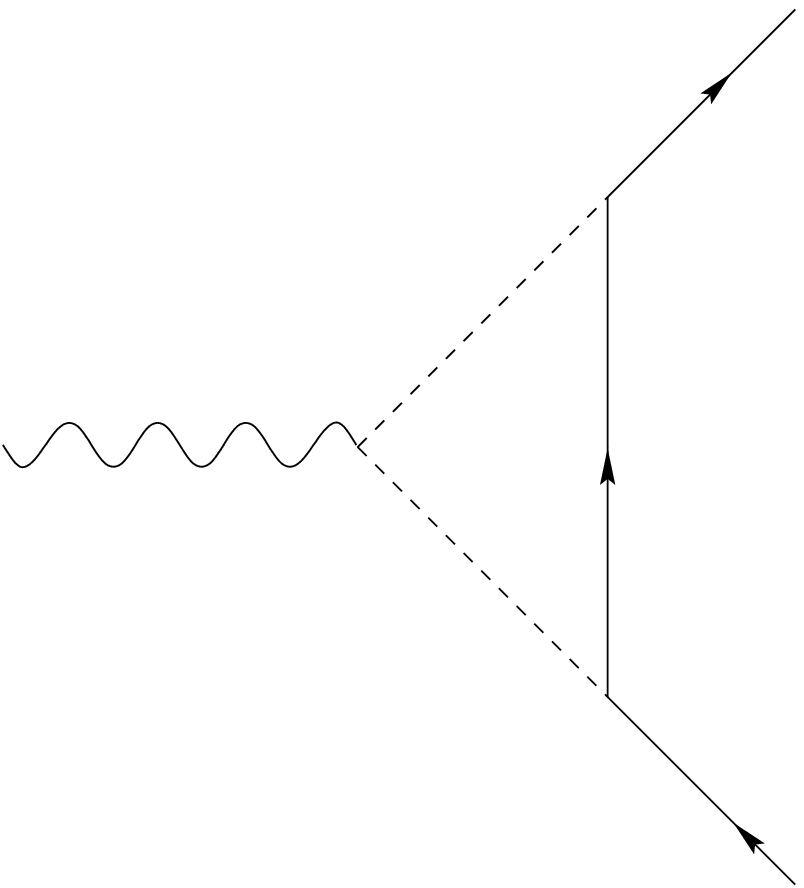,width=3cm}}
\put(10,1.6){$k$}
\put(10.4,3){$p_2$}
\put(10.4,.4){$p_1$}
\put(2.1,1.6){$k$}
\put(2.5,3){$p_2$}
\put(2.5,.4){$p_1$}
\end{picture}
\end{center}
\caption{Vertex diagram in SM (left) and SUSY (right) generating
 a  $\log^2 \frac{s}{M^2}$. $p_1$ and $p_2$ are ingoing. }
\end{figure}
\begin{figure}[htb]\setlength{\unitlength}{1cm}
\begin{center}
\begin{picture}(5,5)
\put(-5,0){\epsfig{file=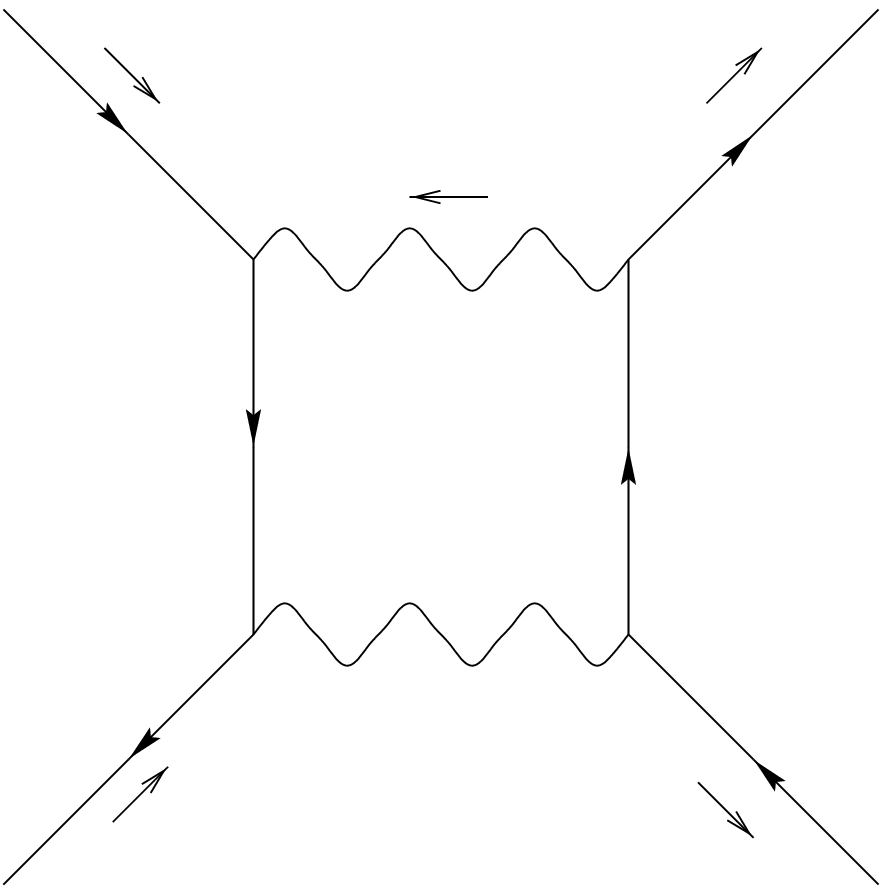,width=3.5cm}}
\put(2,0){\epsfig{file=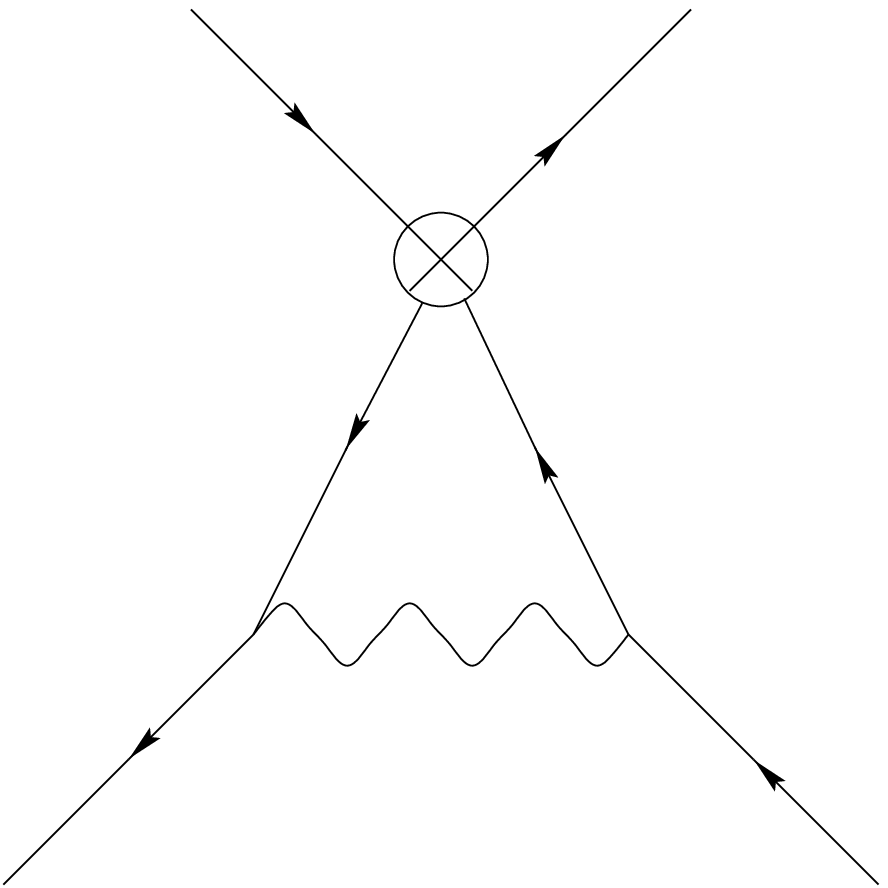,width=3.5cm}}
\put(8,0){\epsfig{file=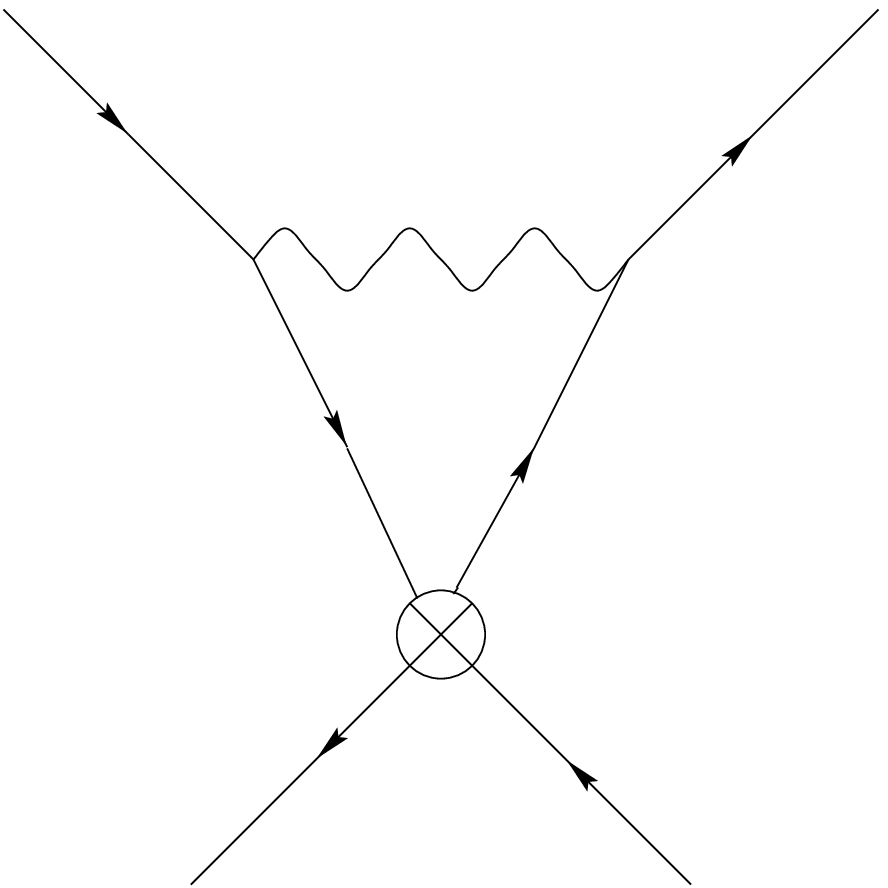,width=3.5cm}}
\put(-5,.7){$p_2$}\put(-5,3){$p_1$}
\put(-1.8,.7){$p_4$}\put(-1.8,3){$p_3$}\put(-3.3,3){$k$}
\put(0,1.7){$\stackrel{IR}{\Longrightarrow}$}
\put(7,1.7){$+$}
\end{picture}
\end{center}
\caption{Box contribution for the SM 
and effective Feynman diagrams in the IR region}
\end{figure}
\begin{figure}[htb]\setlength{\unitlength}{1cm}
\center{\epsfig{file=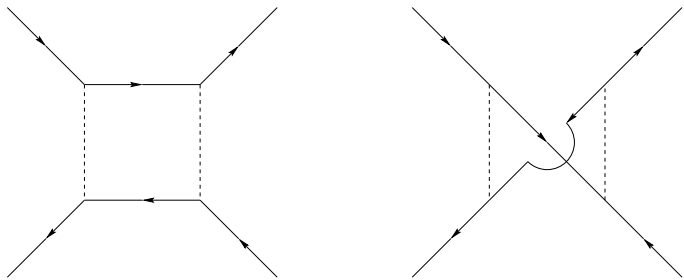,width=9cm}}
\caption{Box contribution for supersymmetry 
(the crossed diagram is also shown)}
\end{figure}
\end{document}